\documentclass[amsmath,amssymb,aps,prd,reprint,twocolumn,showkeys,showpacs]{revtex4-1}

\usepackage[utf8]{inputenc}
\usepackage[utf8]{inputenc}
\usepackage{amsmath}
\usepackage{graphicx}
\usepackage{dcolumn}
\usepackage{bm}
\usepackage{epstopdf}
\printfigures
\usepackage{epsfig}
\usepackage{mathtools}

\usepackage{subfigure}
\usepackage{subfig}

\usepackage{booktabs}

\begin{document}

\title{A note on the dynamical features for the extended $f(P)$ cubic gravity}

\author{Mihai Marciu}
\email{mihai.marciu@drd.unibuc.ro}
\affiliation{Faculty of Physics, University of Bucharest, 405 Atomi\c{s}tilor, POB MG-11, RO-077125, Bucharest-M\u{a}gurele, Romania}

\begin{abstract}
The paper studies the physical characteristics for the extended $f(P)$ cubic gravity from a transitive perspective based on dynamical system analysis, by considering the linear stability theory in two specific cases, corresponding to power--law $f(P)=f_0 P^{\alpha}$  and exponential $f(P)=f_0 e^{\alpha P}$ gravity types, where $f_0$ and $\alpha$ are constant parameters. In these cases we have analyzed the effects in the phase space complexity, revealing the cosmological solutions attached to the critical points. For the power--law and exponential gravity types, we have noticed the presence of two cosmological epochs associated to the critical points involved, corresponding to de--Sitter eras and quintessence--like epochs, described by a constant effective equation of state. For all of these solutions we have studied the dynamical characteristics which are associated to the stability properties, determining possible constraints to various parameters from a transient perspective. The dynamical prospects asserted that the extended $f(P)$ cubic gravity can represent a promising modified theory of gravitation, leading to the manifestation of the accelerated expansion at late time evolution.  
\end{abstract}

\maketitle

\section{Introduction}
\par 
An important topic in the present cosmological context is related to the introduction of various modifications for the gravitational sector, adding to the Einstein--Hilbert action various components based on viable geometrical quantities. The development of such viable theories of gravitation based on different geometrical components can lead to new premises which might explain the current state of the Universe, the past history in the corresponding evolution, bordering the theoretical framework for interesting viewpoints and applications. In the modified gravity context \cite{Nojiri:2017ncd}, an important direction was established with the appearance of the $f(R)$ theory of gravitation \cite{DeFelice:2010aj} which revise the Einstein--Hilbert action by introducing a specific function $f$ based on the scalar curvature $R$. In this manner, this approach paved the way for the development of various theories of gravitation based on different components \cite{CLIFTON20121}. The modified gravity theory based on the $f(R)$ action represent an interesting theory which have been studied in a variety of analyses \cite{Sotiriou:2006hs,Appleby:2007vb,Ananda:2007xh, Amendola:2006kh,Sotiriou:2006qn,Perivolaropoulos:2019vkb}. Furthermore, the introducing of the $f(G)$ theory of gravitation \cite{Nojiri:2005jg,Cognola:2006eg} have occurred naturally within this context, a specific theory based on the Gauss--Bonnet term $G$ which represents an invariant which can lead to interesting cosmological effects \cite{Cognola:2006sp,Guo:2009uk}. Later on, in the recent years, different modified gravity theories have been constructed in the scalar tensor theories based on general relativity and also teleparallel gravity.  The modified gravity theories which add higher--order terms to the Einstein Hilbert action represent viable theoretical approaches which can originate from string theory \cite{GROSS198741}, particular attempts of a more complete and renormalizable \cite{PhysRevD.16.953} theory for the gravitational interaction.
\par 
Within this framework, the Einsteinian cubic gravity \cite{Bueno:2016xff} represents a particular theory for the gravitational interaction which is based on a specific contraction of the Riemann tensor at the cubic order. The introduction in the general framework of modified gravity theories for the Einsteinian cubic gravity has been done recently in a paper \cite{Bueno:2016xff} which introduced the specific form of the corresponding nontopological term $P$ which represents the foundation of the latter theory. The authors have investigated the specific linearization technique in the context of higher-order gravity theories, obtaining the Einsteinian cubic gravity. Furthermore, in the recent years several authors \cite{Bueno:2016lrh,Hennigar:2016gkm,Feng:2017tev,Adami:2017phg,Hennigar:2018hza,Bueno:2018xqc,Poshteh:2018wqy,Jiang:2019kks,Cano:2019ozf,Frassino:2020zuv,Cisterna:2018tgx,Mehdizadeh:2019qvc,Bueno:2019ltp} have investigated different applications of the Einsteinian cubic gravity.

\par 
In a recent paper \cite{Erices:2019mkd} the authors have proposed a new modification of gravity, adding to the Einstein--Hilbert action a viable geometrical model $f(P)$ based on a specific invariant $P$, which encodes specific contractions of the Riemann tensor in the cubic order \cite{Bueno:2016xff}. As shown in the paper, this nontopological term can lead to second order equations in the cosmological context for specific interrelations between various parameters associated to the cubic Riemann component. After extending the action by proposing a new type of gravity type in scalar tensor theories, the authors have investigated the dynamical consequences by adopting a numerical approach based on fine--tuning methods. In the analysis different specific models for the $f(P)$ have been considered, which includes linear, power--law models, and cases where the superposition of the latter cases is exhibited. The study \cite{Erices:2019mkd} revealed that the resulting specific evolution can exhibit quintessence, phantom and quintom behaviors for the early time, while at late time the model evolves near the cosmological constant boundary. Hence, in principle this model can represent a viable theory of gravitation, exhibiting the accelerated expansion as a fundamental dynamical effect. 
\par 
In this context, we have further studied the new gravity type proposed recently by Erices \textit{et al.} \cite{Erices:2019mkd} by considering an approach based on linear stability theory \cite{Bahamonde:2017ize}. The linear stability theory represents a powerful analytical technique which associates a phase space structure to a gravity model constructed in scalar tensor theories, revealing the cosmological epochs in the evolution and some of the viable trajectories which can lead to the present era characterized by the accelerated expansion. In the modified gravity context the linear stability theory \cite{Bahamonde:2017ize} have been considered by various authors in different studies \cite{Alho:2016gzi,Frusciante:2013zop,Gonzalez:2006cj,Guo:2013swa,Shabani:2013mya,Bahamonde:2019urw,Kofinas:2014aka,Jamil:2012nma,Jamil:2012yz,Leon:2014yua,Nersisyan:2016hjh,Roy:2014yta}, revealing possible epochs which can appear in the evolution of the Universe.
\par 
The current paper continues with the latter mentioned technique in modified gravity context, analyzing the possible dynamical effects for the cubic extension of gravity. The plan of the paper is the following. In the Section~\ref{sec:adoua} we present the basic ingredients of the extended cubic gravity and the modified Friedmann relations which are obtained. Furthermore, in Section~\ref{sec:atreia} we investigate the power law type of gravity $f(P)=f_0 P^{\alpha}$, where $f_0$ and $\alpha$ are constant parameters which encode the effects from the geometrical coupling to the new nontopological invariant $P$ based on third order contractions of the Riemann tensor. The Section~\ref{sec:apatra} continues with the exponential gravity type where $f(P)=f_0 e^{\alpha P}$. Finally, in Section~\ref{sec:acincia} we present a short summary and the final concluding remarks for the specific gravity type, the cubic extension of Einstein--Hilbert action.

\section{The equations for the extended $f(P)$ cubic gravity}
\label{sec:adoua}

Within this section we shall present the action and the field equations for the extended $f(P)$ cubic gravity in the case of the FRW cosmological model which assumes the Robertson--Walker metric of the type 
\begin{equation}
\label{metrica}
ds^2=-dt^2+a^2(t) \delta_{ik}dx^i dx^k,
\end{equation}
where $a(t)$ represents the cosmic scale factor, and $\delta_{ik}$ the discrete Kronecker symbol. In this geometrical setup we shall define the Hubble parameter in the usual way as $H=\dot{a}/a$, where the dot(s) represents the derivative with respect to the cosmic time, and $'$ the derivative of a function with respect to its argument. In what follows we rely the presentation of the theoretical arguments for the extended $f(P)$ cubic gravity on different aspects introduced by Erices \textit{et al.} in a recent paper \cite{Erices:2019mkd}. Hence, we consider an extension of the Einstein--Hilbert action by including in the geometrical background the extended $f(P)$ cubic gravity based on the $P$ invariant, a new theory having the following action \cite{Erices:2019mkd}:
\begin{equation}
\label{actiune}
S=S_m+\int d^4x \sqrt{-g} \Bigg( \frac{R}{2}+f(P)\Bigg),
\end{equation}
where $f=f(P)$ is a functional which depends on the nontopological cubic invariant $P$ \cite{Bueno:2016xff} defined as:
\begin{multline}
P=\beta_1 R_{\mu\quad\nu}^{\quad\rho\quad\sigma}R_{ \rho\quad\sigma}^{\quad \gamma\quad\delta}R_{\gamma\quad\delta}^{\quad\mu\quad\nu}+\beta_2 R_{\mu\nu}^{\rho\sigma}R_{\rho\sigma}^{\gamma\delta}R_{\gamma\delta}^{\mu\nu}
\\+\beta_3 R^{\sigma\gamma}R_{\mu\nu\rho\sigma}R_{\quad\quad\gamma}^{\mu\nu\rho}+\beta_4 R R_{\mu\nu\rho\sigma}R^{\mu\nu\rho\sigma}+\beta_5 R_{\mu\nu\rho\sigma}R^{\mu\rho}R^{\nu\sigma}
\\+\beta_6 R_{\mu}^{\nu}R_{\nu}^{\rho}R_{\rho}^{\mu}+\beta_7 R_{\mu\nu}R^{\mu\nu}R+\beta_8 R^3,
\end{multline}
with $\beta_i,i=\overline{1,8}$ constant parameters. The action for the matter component is denoted as $S_m$, encoding the physical effects of a perfect fluid having a barotropic equation of state $p_m=w_m\rho_m$, with $p_m$ the pressure, $\rho_m$ the density, and $w_m$ the state parameter. As can be noted, the nontopological cubic invariant $P$ is based on different contractions of the Riemann tensor in the third order, a geometrical construction which lead to second order field equations if the following conditions for the constant parameters are imposed \cite{Erices:2019mkd, Bueno:2016xff}:
\begin{equation}
\beta_7=\frac{1}{12}\big[3\beta_1-24\beta_2-16\beta_3-48\beta_4-5\beta_5-9\beta_6\big],
\end{equation}
\begin{equation}
\beta_8=\frac{1}{72}\big[-6\beta_1+36\beta_2+22\beta_3+64\beta_4+5\beta_5+9\beta_6\big],
\end{equation}
\begin{equation}
\beta_6=4\beta_2+2\beta_3+8\beta_4+\beta_5.
\end{equation}
Furthermore, let us define an additional parameter $\bar{\beta}$ based on the following identity:
\begin{equation}
\bar{\beta}=(-\beta_1+4\beta_2+2\beta_3+8\beta_4).
\end{equation}
It can be shown that using these parameter's constraints for the FRW metric \eqref{metrica}, the nontopological cubic invariant $P$ reduces to:
\begin{equation}
\label{PP}
P=6\bar{\beta}H^4 (2H^2+3\dot{H}),
\end{equation}
containing derivatives of the cosmic scale factor in the second order. 
\par
For the specific theory characterized by the action \eqref{actiune} we can obtain the modified Friedmann relations by the principle of least action in the case of FRW metric:
\begin{equation}
3H^2=\rho_m+\rho_{de},
\end{equation}
\begin{equation}
3H^2+2\dot{H}=-p_m-p_{de}.
\end{equation}

\par 
In this manner we can define the energy density of the geometrical contribution of dark energy \cite{Erices:2019mkd}, 
\begin{equation}
\rho_{de}=-f(P)-18\bar{\beta} H^4\bigg[H\frac{d}{dt}-H^2-\dot{H}\bigg]f'(P),
\end{equation}
its corresponding pressure component,
\begin{multline}
p_{de}=f(P)+6\bar{\beta} H^3\bigg[H\frac{d^2}{dt^2}+2(H^2+2\dot{H})\frac{d}{dt}
\\-3H^3-5H\dot{H}\bigg]f'(P),
\end{multline}
the expressions for the dark energy equation of state
\begin{equation}
w_{\bf{de}}=\frac{p_{de}}{\rho_{de}},
\end{equation}
and the effective equation of state associated to the cosmological model:
\begin{equation}
w_{\bf{eff}}=\frac{p_m+p_{de}}{\rho_{m}+\rho_{de}}.
\end{equation}
Note that in the subsequent calculations we shall omit the bar in the notation for the $\bar{\beta}$ constant parameter, considering $\beta \vcentcolon = \bar{\beta}$. Since the matter sector is decoupled from the geometrical component, the dark energy sector satisfies a continuity equation, 
\begin{equation}
\dot{\rho_{de}}+3H(\rho_{de}+p_{de})=0
\end{equation}
and the matter constituent satisfies a similar relation,
\begin{equation}
\dot{\rho_{m}}+3H(\rho_{m}+p_{m})=0.
\end{equation}

\par 
Furthermore, let us define the density parameter for the matter component 
\begin{equation}
\Omega_m=\frac{\rho_{m}}{3H^2},
\end{equation}
and the corresponding density parameter associated to the geometrical dark energy component
\begin{equation}
\Omega_{de}=\frac{\rho_{de}}{3H^2},
\end{equation}
yielding the usual constraint
\begin{equation}
\Omega_m+\Omega_{de}=1.
\end{equation}

\section{The power law $f(P)$ cubic gravity}
\label{sec:atreia} 
\par 
In this section we shall study the physical features of the $f(P)$ cubic gravity by considering the linear stability theory, in the case of a power law decomposition $f(P)=f_0 P^{\alpha}$, where $f_0$ and $\alpha$ are constant parameters. We proceed by introducing the following auxiliary variables which permits us to close the dynamical system and approximate the evolution in the first order as an autonomous system of ordinary differential equations:
\begin{equation}
x_1=\frac{\rho_m}{3H^2},
\end{equation}
\begin{equation}
x_2=\frac{f(P)}{3H^2},
\end{equation}
\begin{equation}
x_3=6 \beta H^3 \frac{d^2 f(P)}{dP^2}\dot{P}=6 \beta H^3 \alpha (\alpha-1) \frac{f(P)}{P^2} \dot{P},
\end{equation}
\begin{equation}
x_4=2 \beta H^4 \frac{d f(P)}{dP}=2 \beta H^4 \alpha \frac{f(P)}{P}.
\end{equation}
Furthermore, we consider the transformation of the cosmological system from the cosmic time $t$ to $N$, where $N$ represents the logarithm of the comic scale factor, $N=log(a)$. In these variables, the Friedmann constraint equation have the following form:
\begin{equation}
x_1=-\alpha  x_2+x_2+x_3-x_4+1,
\end{equation} 
while the second Friedmann relation, the acceleration equation can be written as:
\onecolumngrid
\begin{equation}
\label{EEQ2}
\ddot{P}=\frac{H^2 P \left(-(\alpha -1) x_4 \left(9 x_1 w_m+3 x_4+5\right)-(\alpha -2) x_3^2+2 (\alpha -1) x_4 x_3-(\alpha -1) x_2 \left(2 \alpha +4 \alpha  x_3+(9-15 \alpha ) x_4\right)\right)}{9 (\alpha -1)^2 x_4^2}.
\end{equation} 
\twocolumngrid
From the definition of the cubic term $P$ in eq. \eqref{PP} we can write  the following relation:
\begin{equation}
\dot{H}=\frac{1}{3} \left(\frac{\alpha x_2}{x_4}-2\right).
\end{equation}
Moreover, the effective (total) equation of state can be written as:
\begin{equation}
w_{\bf{eff}}=-\frac{2 \alpha  x_2}{9 x_4}-\frac{5}{9}.
\end{equation}
The evolution associated to the cosmological model can be approximated as an autonomous first order system of differential equations where only three auxiliary variables are independent ($x_2, x_3, x_4$),
\onecolumngrid 
\begin{equation}
\label{ua1}
\frac{dx_2}{dN}=\frac{x_2 \left(-2 (\alpha -1) \alpha  x_2+\alpha  x_3+4 (\alpha -1) x_4\right)}{3 (\alpha -1) x_4},
\end{equation}
\begin{equation}
\frac{dx_3}{dN}=3 (\alpha -1) x_4 \mho +\frac{(\alpha -2) x_3^2}{3 (\alpha -1) x_4}+x_3 \left(\frac{\alpha  x_2}{x_4}-2\right),
\end{equation}
\begin{equation}
\label{ua2}
\frac{dx_4}{dN}=\frac{1}{3} \left(4 \alpha  x_2+x_3-8 x_4\right),
\end{equation}
where 
\begin{equation}
\mho=\frac{\ddot{P}}{H^2 P}.
\end{equation}
\twocolumngrid
Considering the acceleration relation \eqref{EEQ2} in terms of the auxiliary variables, we can obtain the final form specific for the variation of $x_3$:
\onecolumngrid 
\begin{equation}
\frac{dx_3}{dN}=\frac{1}{3} \left(x_2 \left(15 \alpha +9 (\alpha -1) w_m-\frac{\alpha  \left(x_3+2\right)}{x_4}-9\right)+9 x_4 w_m-x_3 \left(9 w_m+4\right)-9 w_m-3 x_4-5\right).
\end{equation} 
\twocolumngrid
For the power law $f(P)$ cubic gravity, where $f(P)=f_0 P^{\alpha}$ with $f_0, \alpha$ constant parameters we have determined the critical points of the autonomous system by analyzing the case where the r.h.s. of the eqs. \eqref{ua1}--\eqref{ua2} are equal to zero. In this specific case the investigation revealed that we have two classes of critical points.  
\par
\begin{figure*}[ht!]
                \includegraphics[height=.2\textwidth]{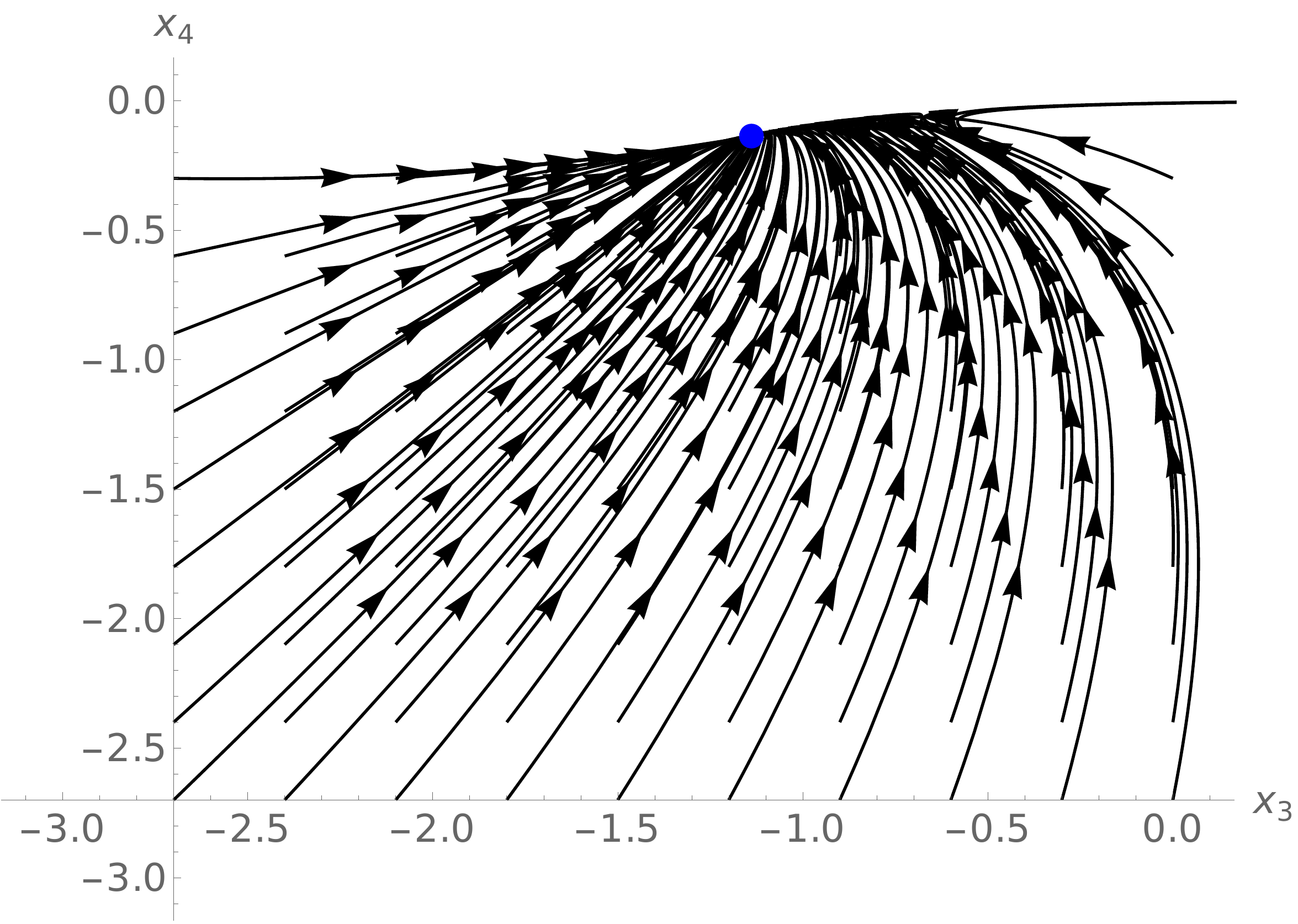}
                \includegraphics[height=.2\textwidth]{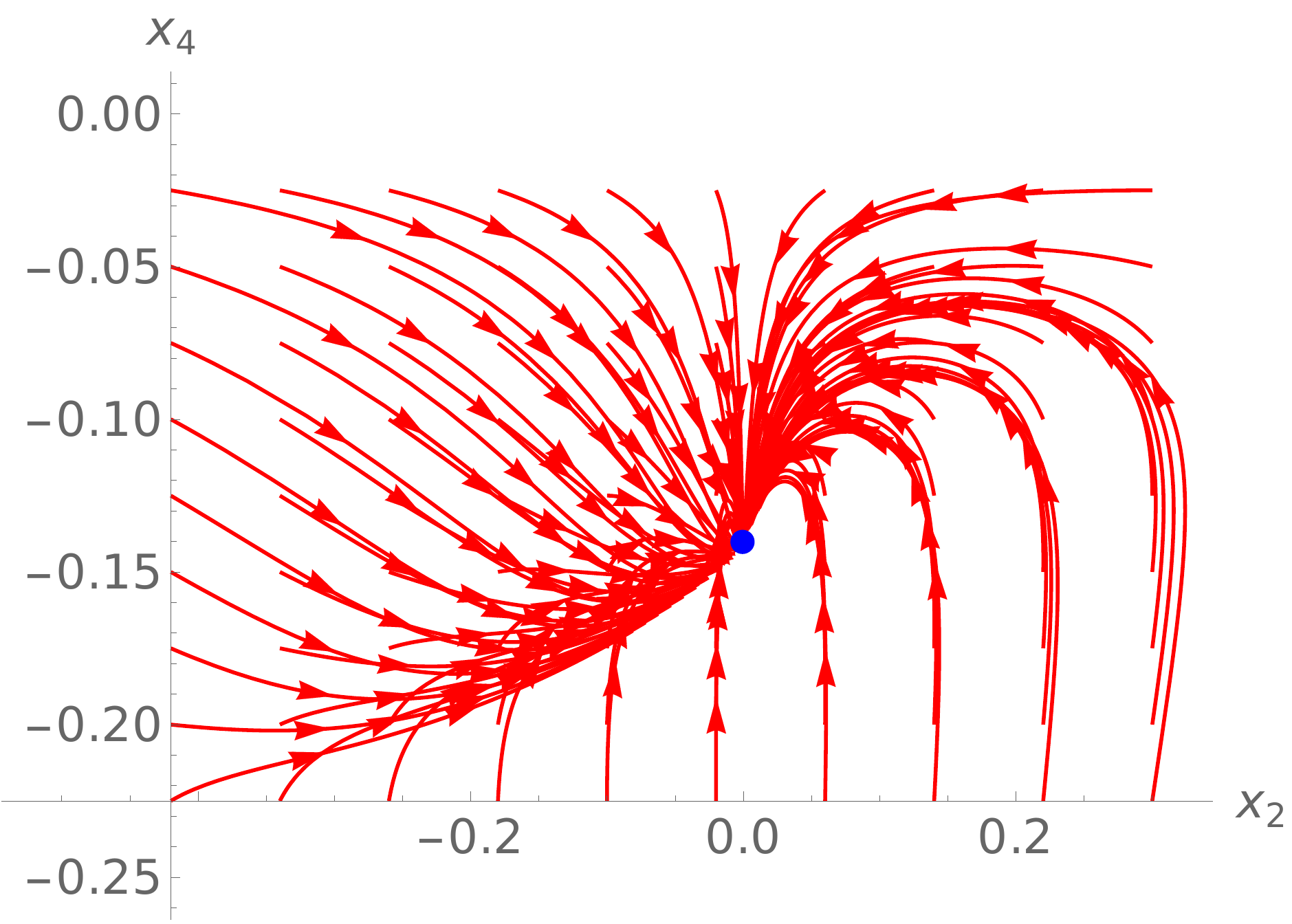}
                \includegraphics[height=.2\textwidth]{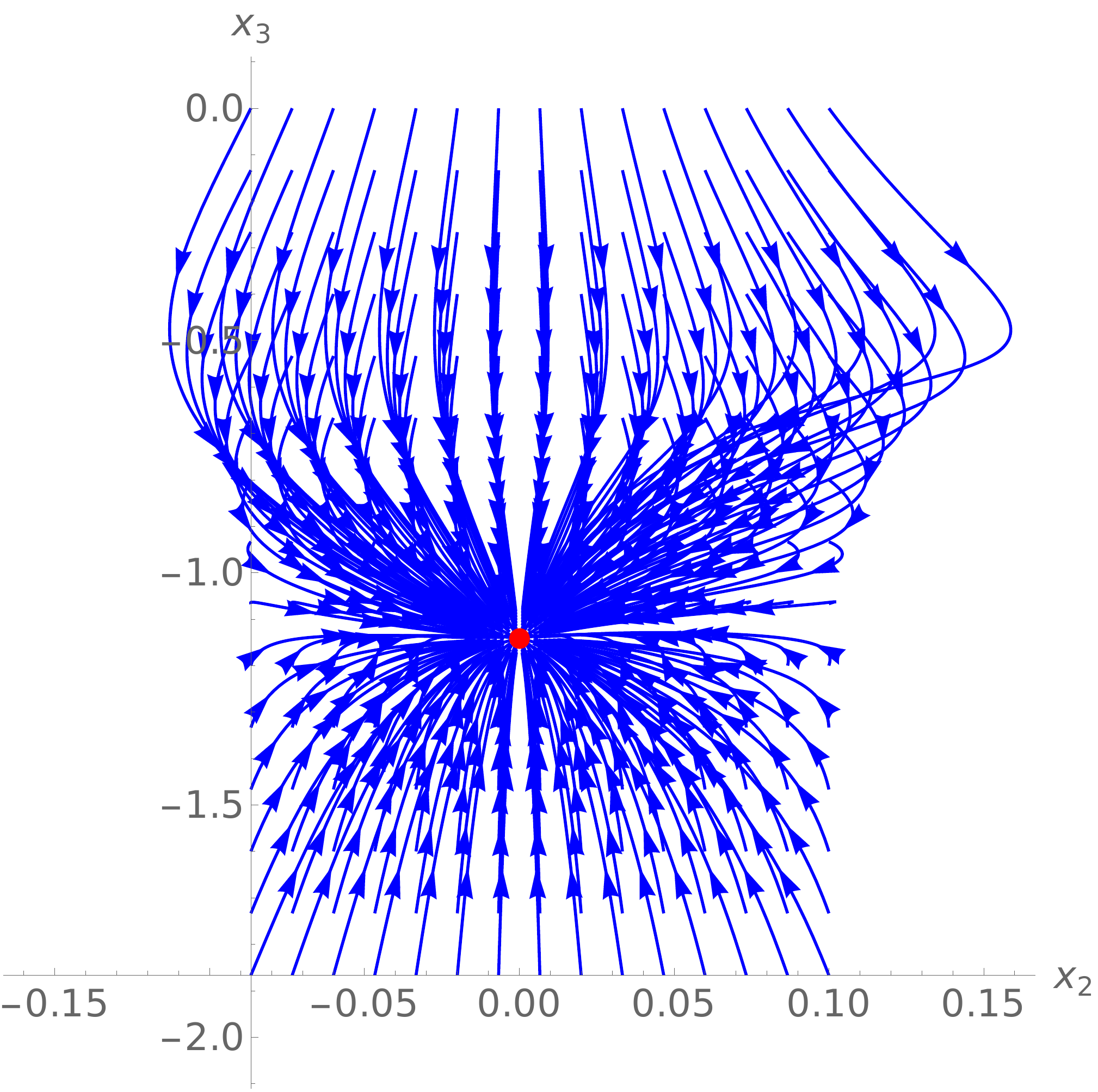}
                \caption{The numerical evolution towards the critical points $P_1$ for specific initial conditions corresponding to a stable scenario ($\alpha=0.5$). }
                \label{fig:fig1}
\end{figure*}

 
The first critical point is located at the following coordinates
\begin{equation}
P_1=\left\{x_2\to 0,x_3\to -\frac{8}{7},x_4\to -\frac{1}{7}\right\},
\end{equation}
representing a cosmological epoch where the dark energy component dominates over the matter sector ($x_1=0$), corresponding to an era associated to a constant equation of state $w_{\bf{eff}}=-\frac{5}{9}$, the geometrical dark energy components acts as a quintessence model. From a dynamical point of view we have obtained the following eigenvalues:
\begin{equation}
E_{P_1}=\bigg[-\frac{7}{3},\frac{4-12 \alpha }{3-3 \alpha },-3 w_m-\frac{5}{3}\bigg].
\end{equation} 
In this epoch, the validation of the following conditions
\begin{equation}
\frac{1}{3}<\alpha <1\land w_m>-\frac{5}{9}
\end{equation}
results in a stable critical point with a quintessential origin where the dark energy phenomenon appears as a geometrical consequence. The evolution in the phase space structure towards the $P_1$ critical point is represented in Fig.~\ref{fig:fig1} for specific initial conditions which are fine--tuned.
\par
\begin{figure*}[ht!]
                \includegraphics[width=.3\textwidth]{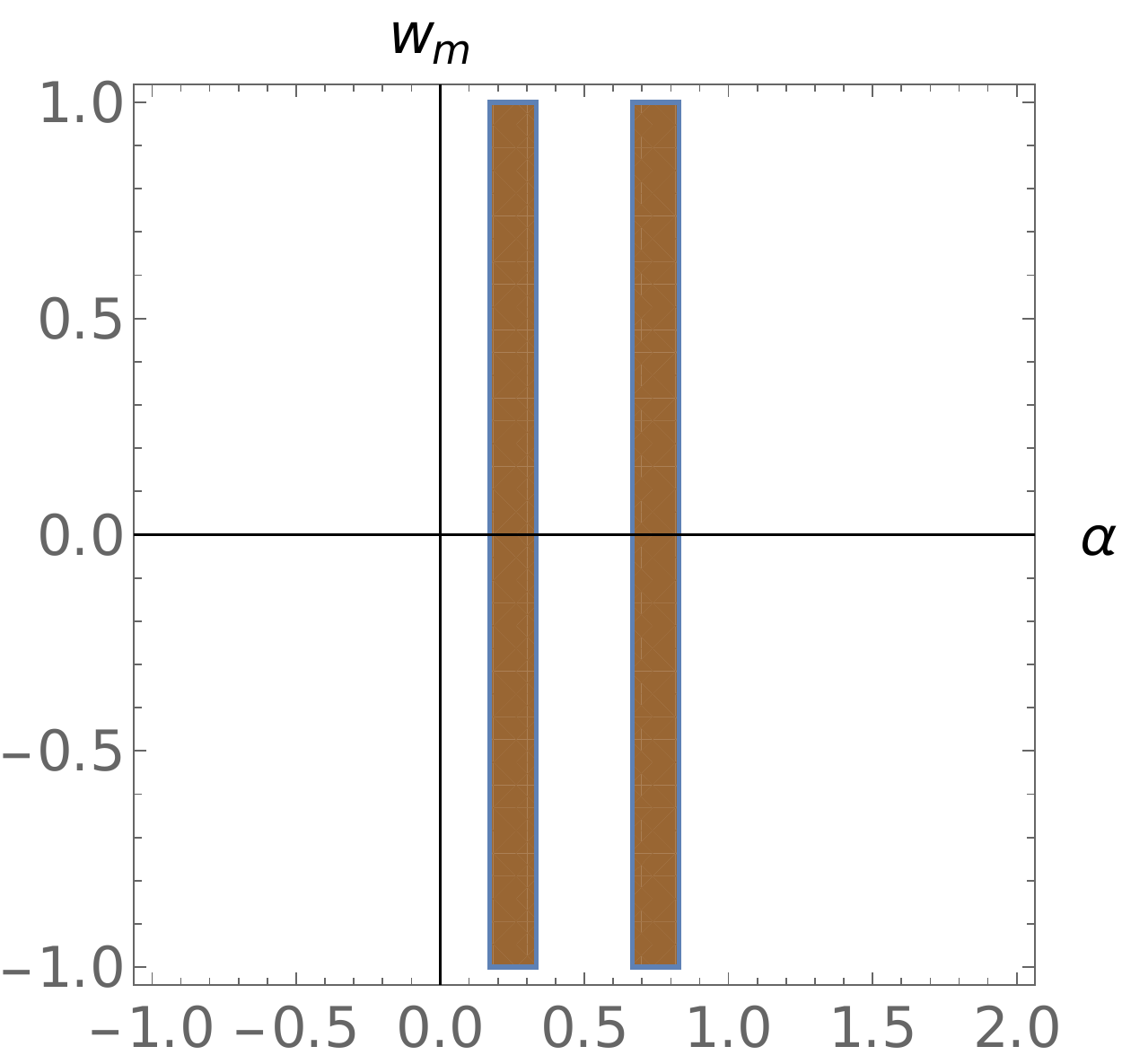}
                \caption{Possible regions where the critical point $P_2$ for the power law scenario $f(P)=f_0 P^{\alpha}$ represents a stable cosmological solution.}
                \label{fig:fig2}
\end{figure*}

\begin{figure*}[ht!]
                \includegraphics[width=.3\textwidth]{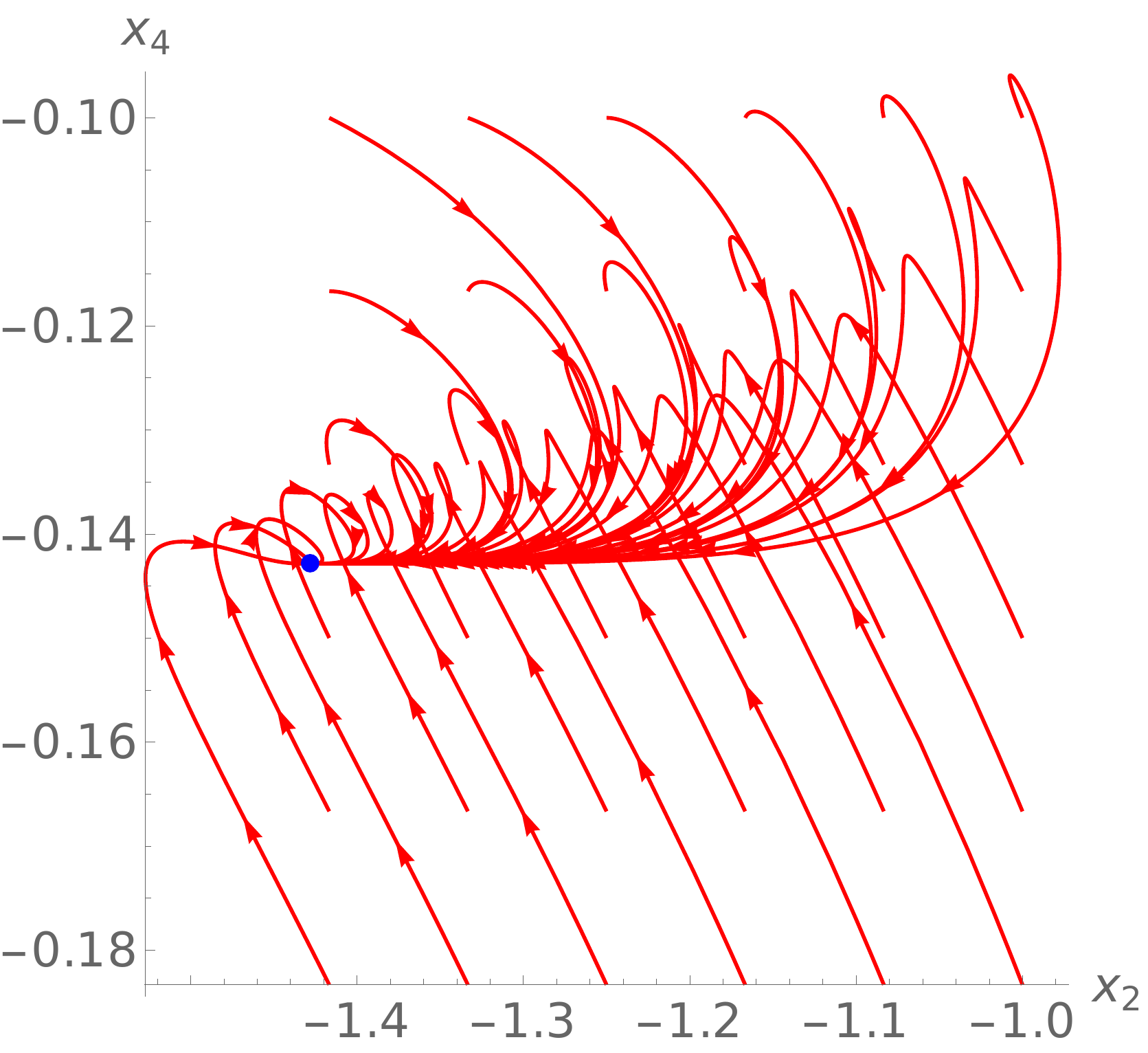}
                \includegraphics[width=.3\textwidth]{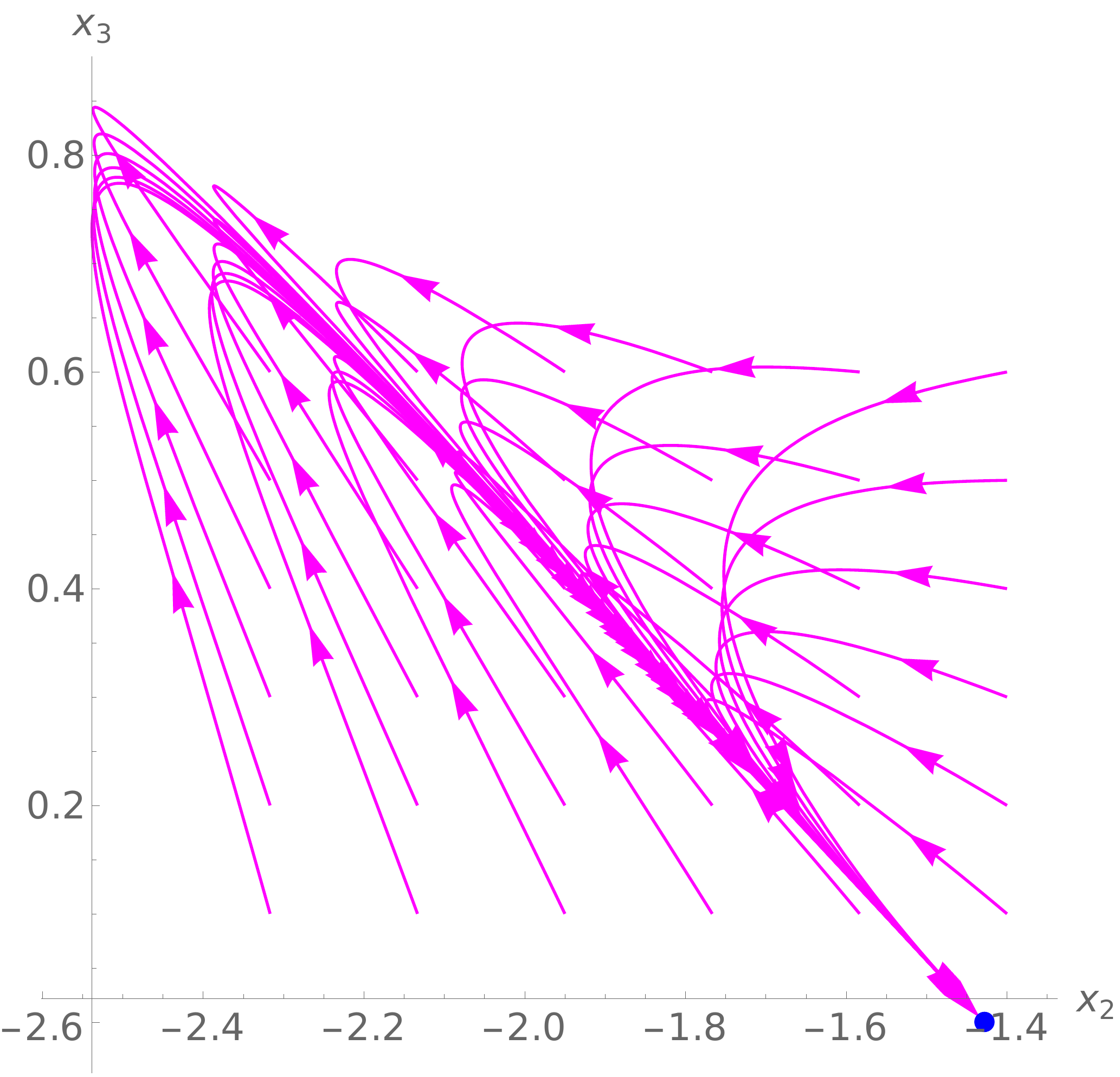}
                \includegraphics[width=.3\textwidth]{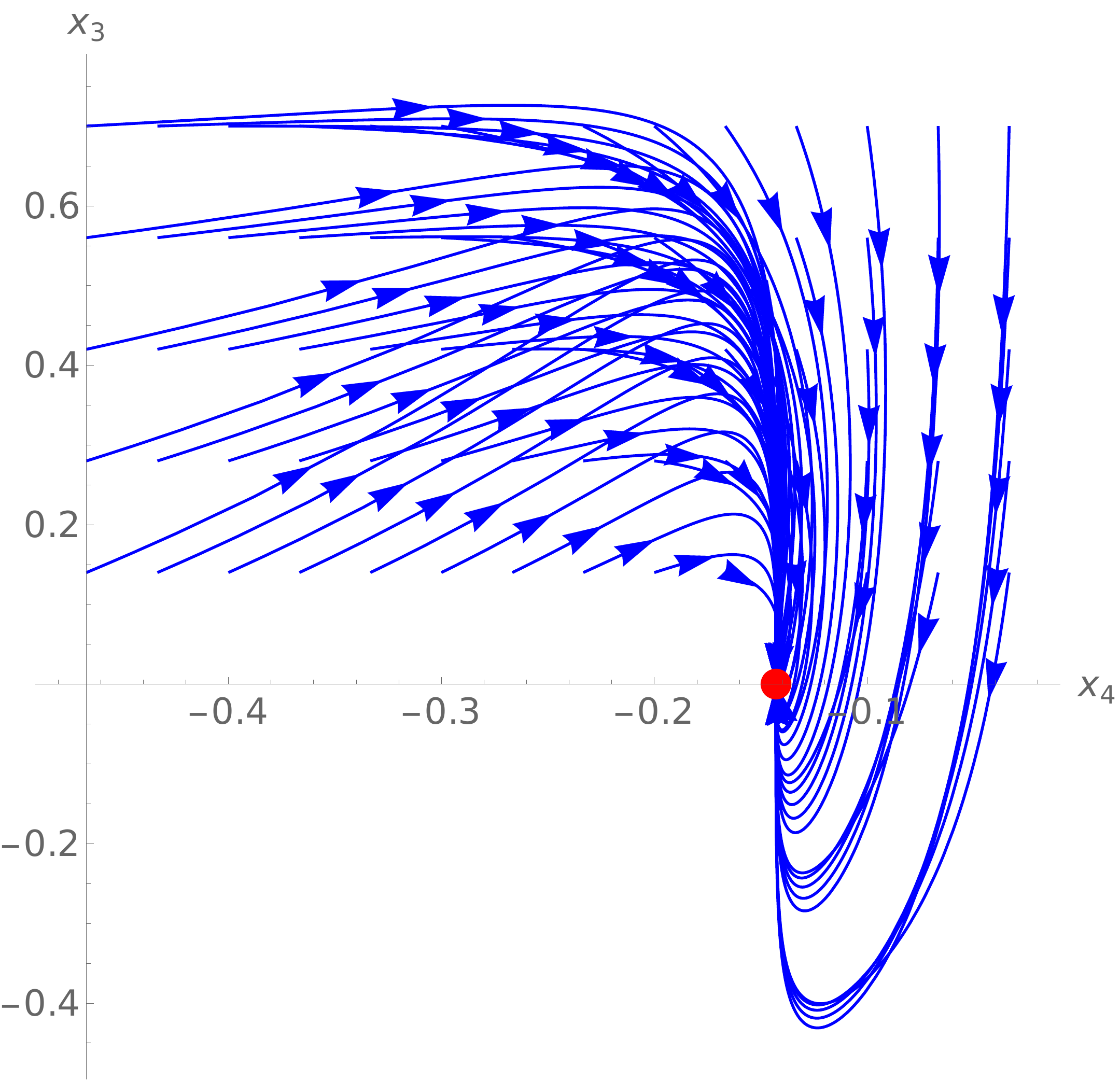}
                \caption{The numerical evolution towards the critical points $P_2$ for specific initial conditions corresponding to a stable scenario ($\alpha=0.2$). }
                \label{fig:fig3}
\end{figure*}

\begin{figure*}[ht!]
                \includegraphics[height=.2\textwidth]{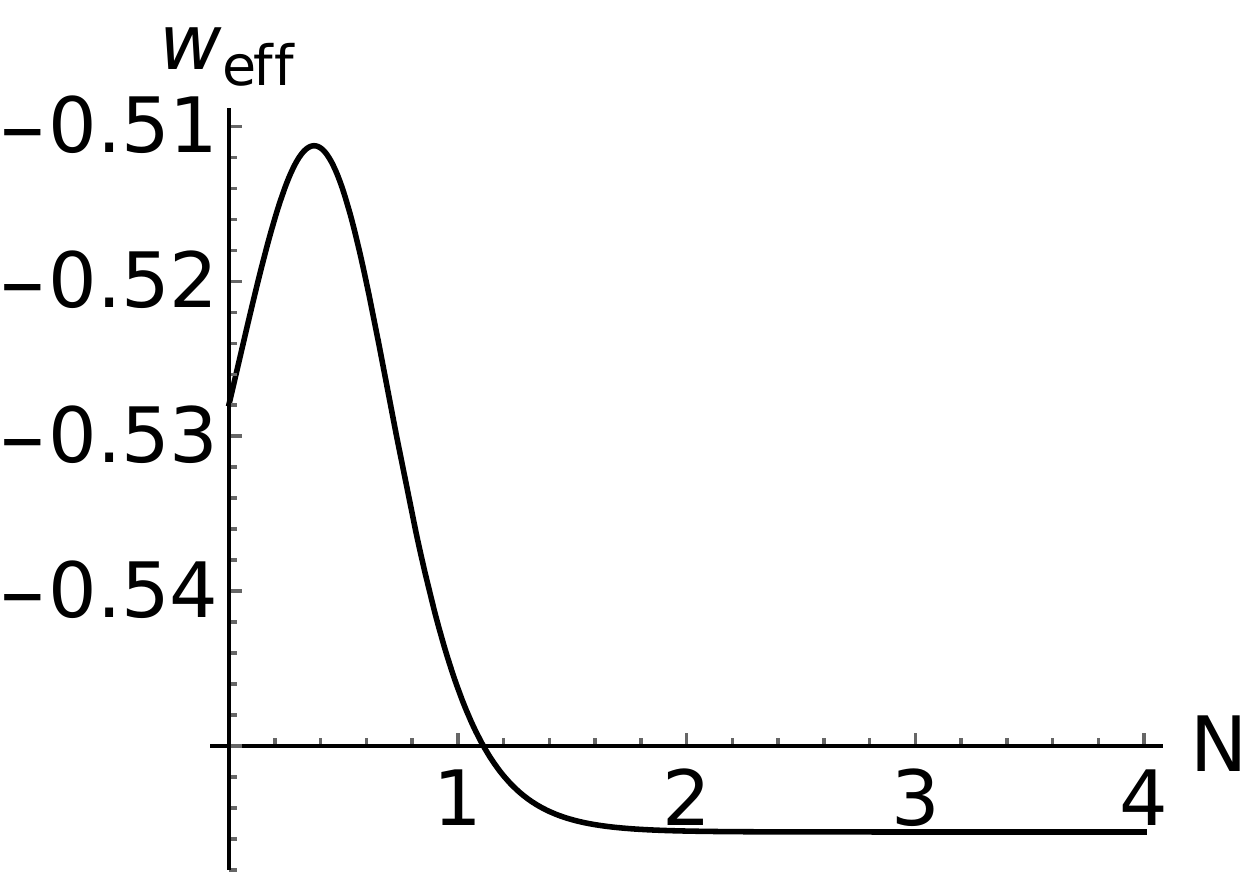}
                \includegraphics[height=.2\textwidth]{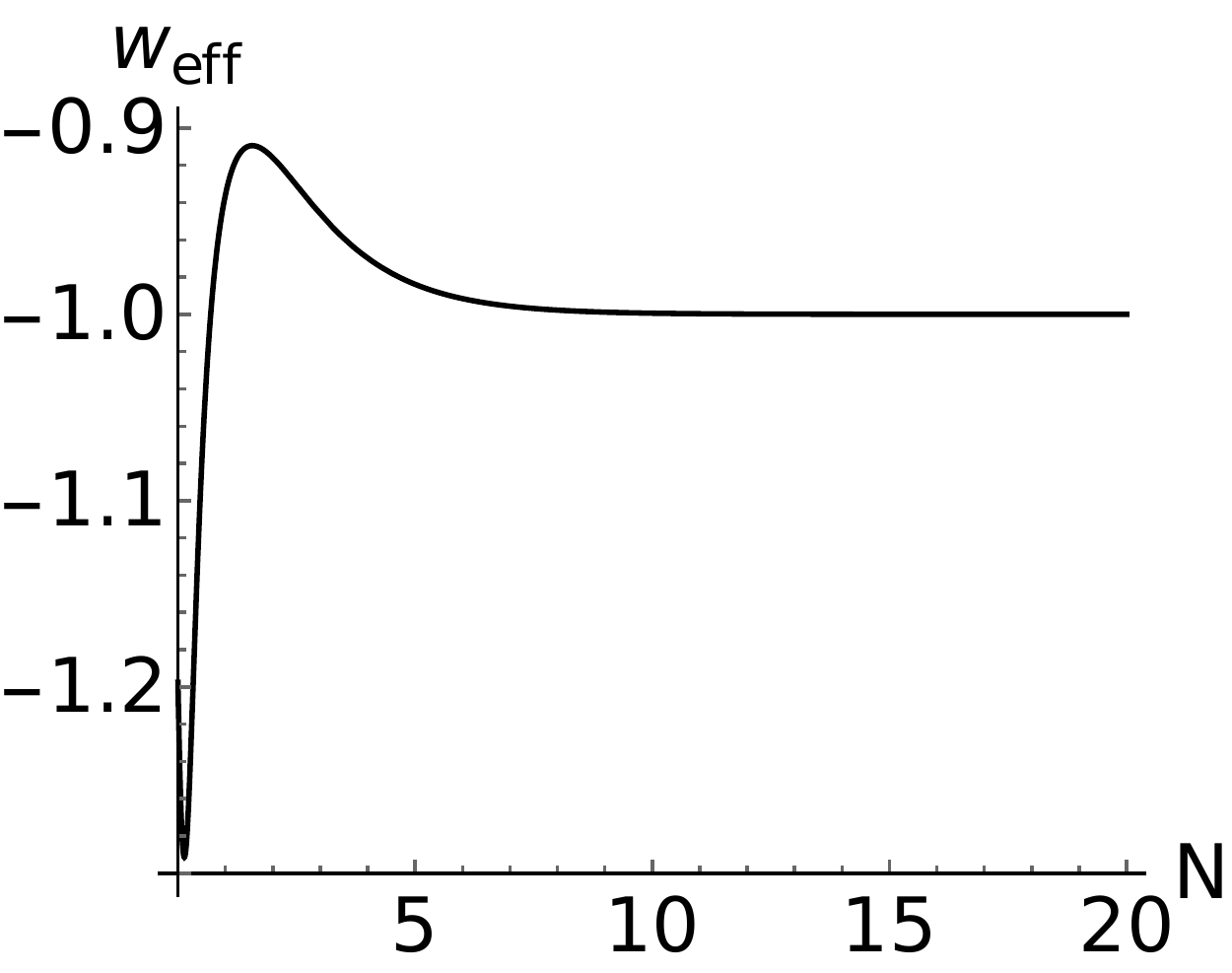}
                \caption{The variation of the effective equation of state $w_{\bf{eff}}$ towards the critical points $P_1$ (left panel) and $P_2$ (right panel) for specific initial conditions corresponding to a stable scenario. }
                \label{fig:fig4}
\end{figure*}

 
The second critical point denoted as $P_2$ is located at the coordinates
\begin{equation}
P_2=\left\{x_2\to \frac{2}{3 \alpha -2},x_3\to 0,x_4\to \frac{\alpha }{3 \alpha -2}\right\},
\end{equation}
describing a de--Siter cosmological epoch where the dark energy component dominates in terms of density parameters, ($x_1=0$). In this case we note that the value of the constant parameters affects the physical location in the phase space structure. The effective equation of state corresponds to a cosmological constant ($w_{\bf{eff}}=-1$), a cosmological solution capable of explaining the current accelerated expansion, a physical effect as a geometrical consequence. The corresponding eigenvalues have the following form:
\onecolumngrid
\begin{equation}
E_{P_2}=\bigg[-3 w_m-3,\frac{\sqrt{225 \alpha ^3-450 \alpha ^2+257 \alpha -32}}{(6-6 \alpha ) \sqrt{\alpha }}+\frac{9 \alpha }{6-6 \alpha }-\frac{9}{6-6 \alpha },-\frac{\sqrt{225 \alpha ^3-450 \alpha ^2+257 \alpha -32}}{(6-6 \alpha ) \sqrt{\alpha }}+\frac{9 \alpha }{6-6 \alpha }-\frac{9}{6-6 \alpha }\bigg].
\end{equation} 
For this specific critical point the validation of the following condition 
\begin{equation}
\left(\frac{1}{30} \left(15-\sqrt{97}\right)<\alpha <\frac{1}{3}\lor \frac{2}{3}<\alpha <\frac{1}{30} \left(\sqrt{97}+15\right)\right)\land w_m>-1
\end{equation}
\twocolumngrid
corresponds to a stable solution, displayed in Fig.~\ref{fig:fig2}. As in the previous case, this cosmological solution  corresponds to the domination of the geometrical dark energy component. The nature of this epoch is affected by the $f(P)$ function and its first variation with respect to the nontopological $P$ component. The evolution in the phase space structure for different initial conditions near the $P_2$ critical point is displayed in Fig.~\ref{fig:fig3}, for specific initial conditions which are fine--tuned. Lastly, we have displayed in Fig.~\ref{fig:fig4} the evolution of the effective equation of state $w_{\bf{eff}}$ towards the critical points $P_1$ and $P_2$ for specific initial conditions corresponding to a stable scenario in both cases, validating the previously obtained results in an analytical manner.

\begin{figure*}[ht!]
                \includegraphics[width=.4\textwidth]{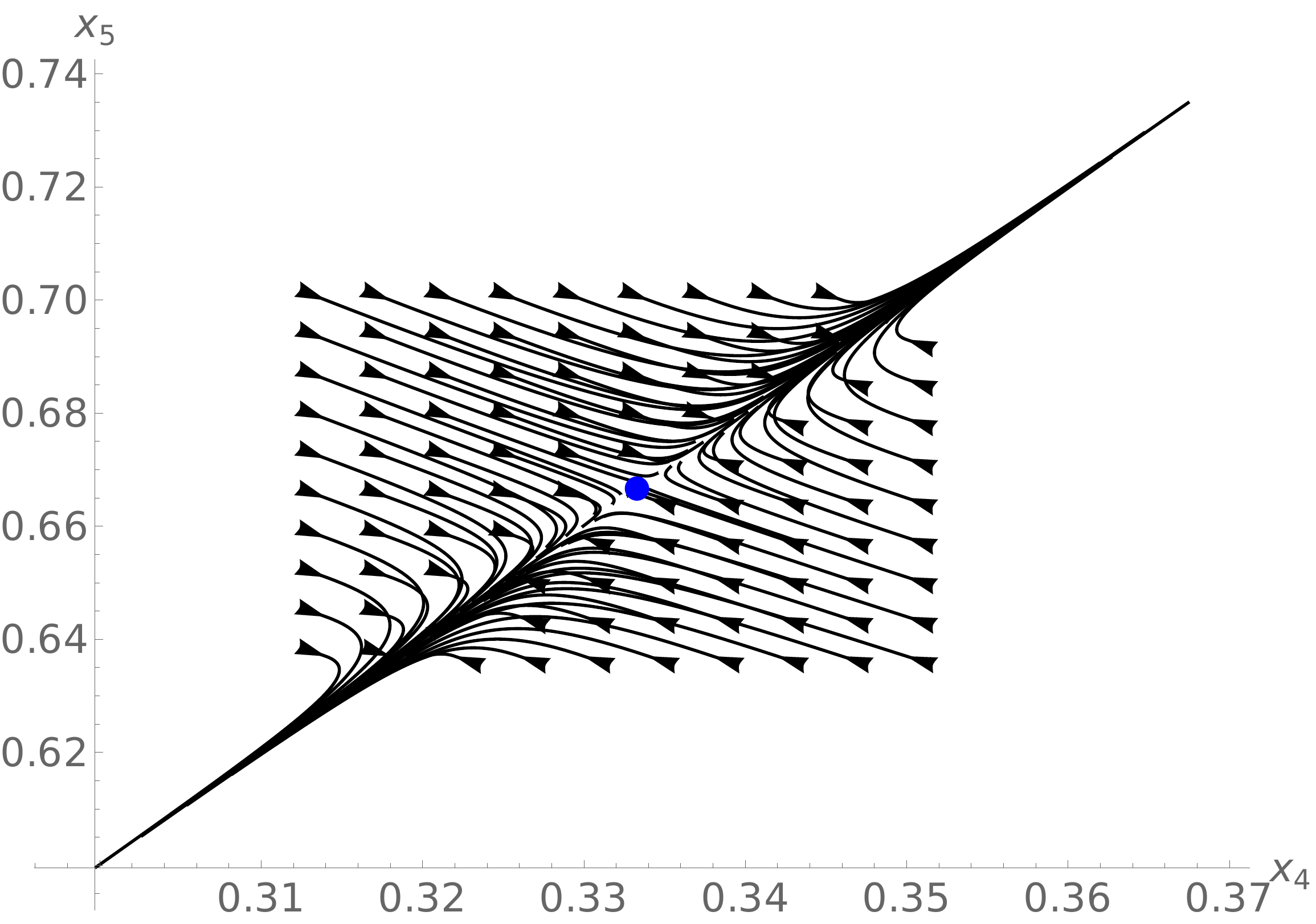}
                \caption{The evolution near the $A_3$ critical point for the exponential $f(P)=f_0 e^{\alpha P}$ cubic gravity, where $f_0$ and $\alpha$ are constant parameters. Different specific initial conditions have been fine--tuned.}
                \label{fig:figlast}
\end{figure*}

\section{The exponential $f(P)$ cubic gravity}
\label{sec:apatra} 
The case of an exponential $f(P)=f_0 e^{\alpha P}$ cubic gravity is also analyzed, where $f_0$ and $\alpha$ are constant parameters. As in the previous section, we proceed with the introduction of the specific auxiliary variables  for the corresponding exponential model:
\begin{equation}
x_1=\frac{\rho_m}{3H^2},
\end{equation}
\begin{equation}
x_2=\frac{f(P)}{3H^2},
\end{equation}
\begin{equation}
x_3=6 \beta H^3 \frac{d^2 f(P)}{dP^2}\dot{P}=6 \beta H^3 \dot{P} \alpha^2 f(P),
\end{equation}
\begin{equation}
x_4=2 \beta H^4 \frac{d f(P)}{dP}=2 \beta H^4 \alpha f(P),
\end{equation}
\begin{equation}
x_5=\frac{d f(P)}{dP} \frac{P}{3 H^2}=\alpha f(P) \frac{P}{3 H^2}.
\end{equation}
In terms of auxiliary variables, the first Friedmann equation (constraint eq.) becomes:
\begin{equation}
x_1=x_2+x_3-x_4-x_5+1,
\end{equation} 
reducing the dimensionality of the autonomous system with one. Furthermore, introducing the specific form of the auxiliary variables into the second Friedmann equation, we get:
\onecolumngrid
\begin{equation}
\ddot{P}=-\frac{H^2 \left(x_4 \left(9 x_1 w_m+9 x_2-15 x_5+5\right)+x_3^2-2 \left(x_4-2 x_5\right) x_3+3 x_4^2+2 x_5\right)}{9 \alpha  x_4^2}
\end{equation}
\twocolumngrid
Considering the transformation from the cosmic time $t$ to $N$, where $N=log(a)$, we can write the evolution of the cosmological model as an autonomous system of ordinary differential equations:
\onecolumngrid
\begin{equation}
\label{exp11}
\frac{dx_2}{dN}=\frac{x_2 \left(x_3+4 x_4-2 x_5\right)}{3 x_4},
\end{equation}
\begin{equation}
\frac{dx_3}{dN}=-\frac{x_4^2 \left(3-9 w_m\right)+x_4 \left(9 \left(x_2+x_3-x_5+1\right) w_m+9 x_2+4 x_3-15 x_5+5\right)+\left(x_3+2\right) x_5}{3 x_4},
\end{equation}
\begin{equation}
\frac{dx_4}{dN}=\frac{1}{3} \left(x_3-8 x_4+4 x_5\right),
\end{equation}
\begin{equation}
\label{exp22}
\frac{dx_5}{dN}=\frac{x_2 x_3+x_5 \left(x_3+4 x_4-2 x_5\right)}{3 x_4}.
\end{equation}
\twocolumngrid
Moreover, we can also write the following relation:
\begin{equation}
\frac{\dot{H}}{H^2}=\frac{\left(x_5-2 x_4\right)}{3 x_4},
\end{equation}
allowing us to determine the value of the effective equation of state for the exponential model, 
\begin{equation}
w_{\bf{eff}}=-\frac{2 x_5}{9 x_4}-\frac{5}{9}.
\end{equation}
The next step in the linear stability theory consists in determining the critical points of the autonomous system \eqref{exp11}--\eqref{exp22} by analyzing the case where the right hand side of the corresponding equations is zero. In this analysis, we have identified three critical points which are subsequently investigated within the present section.
\par 
The first class of cosmological solutions represents a critical surface where the auxiliary variables $x_2$ and $x_5$ are related to $x_4$, corresponding to the first variation of the $f(P)$ function with respect to the $P$ component. The physical location in the four dimensional phase space structure can be written as:
\begin{equation}
A_1=\left\{x_2\to 3 x_4-1,x_3\to 0,x_5\to 2 x_4\right\}.
\end{equation}
These solutions corresponds to a de-Sitter epoch where the total effective equation of state act and describe a cosmological constant behavior, $w_{\bf{eff}}=-1$. At this epoch we also note the domination of the geometrical dark energy component in terms of density parameters, $x_1=0$. The eigenvalues of the Jacobian corresponding to this specific class of cosmological solutions are the following:
\begin{equation}
E_{A_1}=\left\{0,-3(1+w_m),\pm\frac{\sqrt{297 x_4^2-48 x_4+8}}{6 x_4}-\frac{3}{2}\right\},
\end{equation}   
describing a non--hyperbolic solution which cannot be stable due to the behavior of the square root component. Hence, from a dynamical perspective this class of solutions represents a saddle epoch.
\par 
The next critical point $A_2$ is located at the following coordinates:
\begin{equation}
A_2=\left\{x_2\to 0,x_3\to -\frac{8}{7},x_4\to -\frac{1}{7},x_5\to 0\right\},
\end{equation} 
describing an epoch characterized by the domination of the geometrical dark energy component in terms of density parameters ($x_1=0$), having a constant equation of state
\begin{equation}
w_{\bf{eff}}=-\frac{5}{9},
\end{equation}
which corresponds to a quintessence regime. From a dynamical point of view we have obtained the following eigenvalues, 
\begin{equation}
E_{A_2}=\left\{4,4,-\frac{7}{3},-3 w_m-\frac{5}{3}\right\},
\end{equation}
describing a cosmological era which have a saddle comportment which does not depend on the value of the $\alpha$ and $f_0$ constant parameters.
\par 
The last critical point for the $f(P)=f_0 e^{\alpha P}$ cubic gravity case where $f_0$ and $\alpha$ are constant parameters represents also a de--Sitter epoch where
\begin{equation}
A_3=\left\{x_2\to 0,x_3\to 0,x_4\to \frac{1}{3},x_5\to \frac{2}{3}\right\},
\end{equation}
a solution dominated by the non--negligible value of the $x_4$ and $x_5$ variables, embedding physical effects form the specific variation of the geometrical extension in the cubic order. This specific case corresponds to a de--Sitter regime ($w_{\bf{eff}}=-1$) determined by the domination of the geometrical dark energy component ($x_1=0$). The eigenvalues of the Jacobian specific for this type of solutions are the following:
\begin{equation}
E_{A_3}=\{-4,1,0,-3 (w_m+1)\}.
\end{equation}   
Hence the cosmological solution have a saddle dynamical behavior and can explain the current acceleration of the known Universe near the de--Sitter regime. The evolution near the $A_3$ critical point for the exponential $f(P)=f_0 e^{\alpha P}$ cubic gravity have been displayed in Fig.~\ref{fig:figlast} by fine--tuning the initial conditions in the dynamical basin. The analysis presented in this section for the exponential $f(P)$ cubic gravity have shown that the resulting dynamics can exhibit the Universe's acceleration as the specific physical effect, a consequence of the geometrical coupling in the cubic order.

\section{Summary and Conclusions}
\label{sec:acincia} 

In this paper we have studied a recently proposed cosmological theory, the extended $f(P)$ cubic gravity which is based on a novel cubic invariant $P$ which represents a nontopological component leading to second order equations for the gravitational part in the FLRW cosmological background. After presenting the modified Friedmann relations which are obtained by the variation of the action with respect to the inverse metric $g^{\mu\nu}$, we have explored the physical features of the latter theory by adopting the linear stability method. In this case we have considered two specific models, the exponential type and the power law form. In the power--law case the adopted specific functional is described by $f(P)=f_0 P^{\alpha}$, with $f_0$ and $\alpha$ constant parameters, while for the exponential case we have $f(P)=f_0 e^{\alpha P}$, encoding the geometrical aspects of the nontopological cubic invariant $P$. Considering an approach based on the linear stability theory, we have introduced the specific auxiliary variables for each of these models which enables us to approximate the dynamical model as an autonomous system of ordinary differential equations, investigating the locations of the critical points and the corresponding dynamical features encoded into the eigenvalues.
\par 
For these specific models our investigation revealed that the structure of the phase space is not very complex in terms of physical or dynamical features. The cosmological epochs which are associated to the critical points show the existence of two families of dynamical eras, corresponding to different cosmological stages. The first stage represents a de--Sitter epoch where the geometrical dark energy component mimics a cosmological constant solution, a critical point which can explain the present accelerated expansion of the Universe, and the dynamics close to the de--Sitter background. A second stage observed in the analysis is represented by another epoch where the geometrical dark energy component has a constant equation of state. In this epoch, the geometrical dark energy component dominates in terms of density parameters, having a quintessence behavior. In the power law case, the structure of the phase space have three dimensions, associated to the $x_2,x_3,x_4$ auxiliary variables. For this model we have presented the critical points and the corresponding dynamical features in the hyperbolic case, supporting the results also with specific numerical aspects, revealing some trajectories in the phase structure. 
\par 
The second cosmological scenario corresponds to the exponential type $f(P)=f_0 e^{\alpha P}$, where the phase space structure has four independent auxiliary variables, encoded into $x_2, x_3, x_4, x_5$. In this scenario we have obtained three classes of critical points, corresponding to different possible eras for the evolution of the Universe. In this case, we have obtained two critical points which corresponds to a de--Sitter evolution, while the last class corresponds to an epoch characterized by a constant equation of state where the geometrical dark energy acts as a quintessence model. For each critical point obtained we have determined the corresponding eigenvalues which encode the dynamical effects. Hence, in the structure of the phase space the quintessential epoch represents a saddle dynamical behavior, while one of the remaining critical points which corresponds to a de--Sitter evolution is also saddle. The remaining de--Sitter epoch is a critical point which is non--hyperbolic due to the presence of one zero eigenvalue, analyzed only in the saddle context due to the limitations associated to the linear stability theory. 
\par 
For each family of cosmological scenarios investigated, the power--law case and the exponential type, the current analysis revealed that in the phase space structure the cosmological solutions can describe the evolution of the Universe at the background level near the cosmological constant boundary, explaining the accelerated expansion as a fundamental dynamical effect. The existence of these cosmological solutions which can explain the accelerated expansion can solve the dark energy problem as a geometrical effect from the coupling to a nontopological cubic invariant $P$ in the context of the extended $f(P)$ cubic gravity. We can finally note that this gravity type can represent a possible viable theory of gravitation which should be further investigated by adopting various viable approaches.

\section{Acknowledgements}
The development of this project implied the consideration of various computations which have been done in Wolfram Mathematica \cite{Mathematica} using the xAct package \cite{xact}.

\bibliography{bibliografie}
\bibliographystyle{apsrev}

\end{document}